\documentclass[twocolumn]{emulateapj}

\slugcomment{Accepted for publication in \apj}
\shortauthors{Nierenberg et al.}
\shorttitle{Evolution of satellites in cold and warm dark matter}

\newcommand{\be}{\begin{equation}}
\newcommand{\ee}{\end{equation}}
\newcommand{\bea}{\begin{eqnarray}}
\newcommand{\eea}{\end{eqnarray}}

\newcommand{\Mstar}{M^{*}}
\newcommand{\Msun}{M_{\odot}}

\begin{document}

\title{The cosmic evolution of faint satellite galaxies as a test of galaxy formation and the nature of dark matter}

\author{A.M. Nierenberg\altaffilmark{1}, T.~Treu\altaffilmark{1,2}, N. Menci\altaffilmark{3}, Y. Lu\altaffilmark{4}, W. Wang\altaffilmark{5}}
\altaffiltext{1}{Department of Physics, University of California, Santa Barbara, CA 93106, USA; amn01@physics.ucsb.edu}
\altaffiltext{2}{Packard Fellow}
\altaffiltext{3}{NAF - Osservatorio Astronomico di Roma, via di Frascati 33, I-00040 Monteporzio, Italy}
\altaffiltext{4}{Kavli Institute for Particle Astrophysics and Cosmology, Stanford CA, 94309}
\altaffiltext{5}{Key Laboratory for Research in Galaxies and Cosmology of Chinese
Academy of Sciences, Max-Panck-Institute Partner Group, Shanghai
Astronomical Observatory, Nandan Road 80, Shanghai 200030, China }

\begin{abstract}

The standard cosmological model based on cold dark matter (CDM)
predicts a large number of subhalos for each galaxy-size halo.
Matching the subhalos to the observed properties of
luminous satellites of galaxies in the local universe poses a
significant challenge to our understanding of the astrophysics of
galaxy formation. We show that the cosmic evolution and host mass
dependence of the luminosity function of satellites provides a
powerful new diagnostic to disentangle astrophysical effects from
variations in the underlying dark matter mass function.
We illustrate this by comparing recent observations of satellites between  redshifts $0.1<z<0.8$ based on 
Hubble Space Telescope images, with predictions from of three different state-of-the art semi-analytic models applied to CDM 
power spectra, with one model also applied to a Warm Dark Matter (WDM) spectrum. We find that even
though CDM models provide a reasonable fit to the local luminosity
function of satellites around galaxies comparable to the 
Milky Way, they do not reproduce the data as well for
different redshifts and host galaxy stellar masses, indicating that further improvements in the description of star formation are likely needed.
The WDM model matches the observed mass dependence and
redshift evolution of satellite galaxies more closely,
indicating that a modification of the underlying power
spectrum may offer an alternative solution to this tension. 
We conclude by presenting predictions for the color distribution of
satellite galaxies to demonstrate how future observations will be able
to further distinguish between these models and help constrain
baryonic and non-baryonic physics.
\end{abstract}


\section{Introduction}

The standard model based on dark energy and Cold Dark Matter (CDM) is
successful at describing the large-scale ($>$ Mpc) structure of the
universe \citep[e.g.][]{Komatsu++11}.  In this standard model,
galaxies and clusters of galaxies grow hierarchically within dark
matter halos \citep[e.g.,][]{Springel++06}.

At the scales typical of galaxies ($\sim$kpc), the model faces a
number of observational challenges.  A major source of tension between
observation and the standard model is the so-called ``missing
satellite problem'' \citep{Klypin++99, Moore++1999}. At large scales,
a simple ranked matching of simulated subhalos to observed luminous
structure predicts clustering behavior consistent with that of the
observed luminous structure \citep{Conroy++09,Behroozi++10}.  At
smaller scales however, this simple matching scheme breaks down. For
instance, CDM simulations predict that the Milky Way (MW) should have
approximately 10 subhalos with rotation velocities greater than Fornax, 
while only 3 satellites that are more luminous are observed
\citep{Boylan-Kolchin++12,Strigari++12}.  The discrepancy between the
predicted number of subhalos and the observed number of luminous
satellites becomes more dramatic at even lower halo masses \citep{Strigari++07}. 
While the measurement of low luminosity satellites may be biased in the Milky Way due to disk 
obscuration \citep{Toll++08}, the lack of low mass galaxies extends to the field
as well, where \citet{Papastergis++11} measured a factor of 8 fewer
galaxies with velocity widths of $\sim$50 km/s than predicted by current $\Lambda$CDM
simulations. Furthermore, recent surveys have found that there is not
a significant population of optically dark, gas rich galaxies, \citep{Doyle++05}, indicating
that if these very low mass halos exist, they do not contain significant amounts of gas or stars.

The discrepancy may be due to a variety of factors. The complex
physics of star formation in low mass halos makes matching simulated
halos to observed satellite galaxies difficult.  Significant progress
has been made in understanding the complex processes which affect star
formation in low mass halos \citep{Krav++10}.  In particular, numerous
studies have focused on studying how both sub-galactic effects such as
supernovae feedback and stellar winds, and super-galactic effects such
as UV heating from reionization, and tidal and ram pressure stripping
by the central galaxy can suppress star formation and thereby produce
the present day satellite luminosity function
\citep[e.g.][]{Bullock++2000,Benson++2002,Somerville++02,Kravtsov++04,Kaufmann++08,Maccio++10,
Springel++10}. 

In addition to suppressing star formation, thereby making subhalos
undetectable, several studies have examined whether supernovae
feedback can significantly flatten the central regions of dark matter
density profiles. This would potentially explain the discrepancy
between the predicted and observed central velocities of $\Lambda$CDM
subhalos around the Milky Way \citep{Boylan-Kolchin++12}, by lowering
the subhalo central velocities, thereby making them appear less
massive. Simulations with baryons have produced conflicting
predictions as to whether supernovae can
\citep{Governato++12,Teyssier++13} or cannot \citep{Garrison++13}
significantly alter the kinematics in the central regions of dark
matter halos. 

The issue of the low central velocities of Milky Way subhalos is further complicated
by the fact that the $\Lambda$CDM prediction for the subhalo circular velocity function
depends sensitively on the virial mass of the Milky Way halo. \citet{WangJie++12} showed that 
taking into account the uncertainty on the measured Milky Way virial mass, the observed subhalo circular velocity
function was consistent with that observed in ensemble simulations of dark matter halos. \citet{Purcell++12}
further showed that even simulated halos with masses corresponding to the observational mean have subhalo
populations consistent with the Milky Way a significant fraction ($\sim 10 \%$) of the time.

These studies demonstrate
that it is essential to study the statistical properties of satellite galaxies
in observation and simulations for a large sample of host galaxies, in order to determine whether 
apparent discrepancies are driven by the stochastic nature of galaxy formation, or observational uncertainty,
rather than poorly modeled physical processes.

Furthermore, as the baryonic processes which affect the subhalo
population occur over cosmological time
scales, ideally the predicted effects would be compared with
observations of the satellite population over as much of the history
of the universe as possible.

If the solution to this problem is astrophysical in nature, i.e.  low
mass halos exist but do not form stars and are therefore undetected by
traditional astronomical observations, one needs to turn to other
methods to verify their existence. This line of reasoning has
motivated searches for satellites halos based on properties that are
independent of their stellar content, such as their gravitational
lensing effect
\citep[e.g.][]{Mao++98,M+M01,Dalal++02,Ama++06,K+M09,Treu++10,Vegetti++12},
or on their expected DM annihilation signal \citep{Kuhlen++08,
Porter++11, Strigari++12B,A+K12,Fadley++12}, or their influence of tidal streams
in the Milky Way \citep{Carlberg++12}.  Direct detection of dark
subhalos would be a stunning confirmation of the standard model.

However, this is not the only possible solution to the missing
satellite problem.  Warm Dark Matter (WDM)~ \citep[e.g.][and
references therein]{Colombi++96} is an interesting alternative to CDM,
potentially offering an elegant astroparticle solution to the missing
satellite problem \citep{Lovell++12}. In WDM scenarios, small-scale
structure is suppressed relative to CDM. There are a variety of
mechanisms which can achieve this, for instance either by reducing the
mass of the dark matter particle for a thermal relic
\citep[e.g][]{Steffen++06} or by introducing non-thermal particles
such as sterile neutrinos produced from oscillations of active
neutrinos \citep[e.g.][]{Olive++82, Shi++99, Abazajian++01,
Dolgov++02}.  The suppression of small-scale structure in turn affects
the full merger history of a halo and its subhalos, which recent
cosmological simulations have begun to study in detail
\citep{Menci++12,Lovell++12, Kamada++13, Polisensky++11,Bode++01,Gottloeber++10, deVega++12}.

Clearly the astrophysical and astroparticle solutions to the missing
satellite problem are not mutually exclusive, and progress on both
fronts may be needed to reconcile the model with the data. In fact, a
proper observational test of the WDM model predictions for the number
of satellite galaxies requires an accurate description of the
astrophysical effects related to star formation.

One difficulty in studying this issue, is that is impossible to distinguish the effects of
varying baryonic physics from suppressing the subhalo mass function using only
low redshift measurements of the satellite or field luminosity function. However, with multiple
observables, one may begin to disentangle the effects. For instance, \citet{Kang++12},
showed that if the WDM particle is too light, one cannot simultaneously reproduce the Tully-Fisher 
relation and the field luminosity function.

In this paper we present a new test of three semi-analytic galaxy
formation models, with one model implemented for both a Warm and Cold
Dark Matter substructure mass function, in order to show how 
variations in the halo mass function produce different predictions from variations
in baryonic physics. We compare the predictions
from these models with the observed abundance of satellite galaxies as
a function of host galaxy mass and cosmic time. 

This test is made possible by two recent developments.  On the
observational side, the implementation of powerful statistical tools
to detect and count satellites in deep archival Hubble Space Telescope
images provides data to compare with model predictions
\citep{Nierenberg++11, Nierenberg++12}. On the theoretical side,
advances in computational methods have allowed for cosmological
simulations with unprecedented volume and resolution. In this work we
focus on predictions from four independent cosmological simulations,
which have semi-analytic models applied to dark matter merging trees,
which we describe in more detail below.

The paper is organized as follows: In \S \ref{sec:observations} we
describe our observations of the satellite luminosity function. In \S
\ref{sec:theory} we summarize the key aspects of the theoretical
models. In \S \ref{sec:results}, we compare the observations with the
theoretical predictions. In \S \ref{sec:comparison of sims} we compare
a few of the main properties of the models and present new predictions
for the distribution of colors of satellite galaxies across cosmic
time.  In \S \ref{sec:literature} we compare the model results with
other observations taken from the literature.  Finally in \S
\ref{sec:discussion} we conclude with a discussion and summary of the
results.

\section{Observations}
\label{sec:observations}

Measurements of the satellite population were obtained by \citet{Nierenberg++12},
using ACS F814W imaging of the COSMOS field, where a full description of the analysis is given. 

\begin{figure*}
\begin{center}
\includegraphics[width=0.98\textwidth,clip]{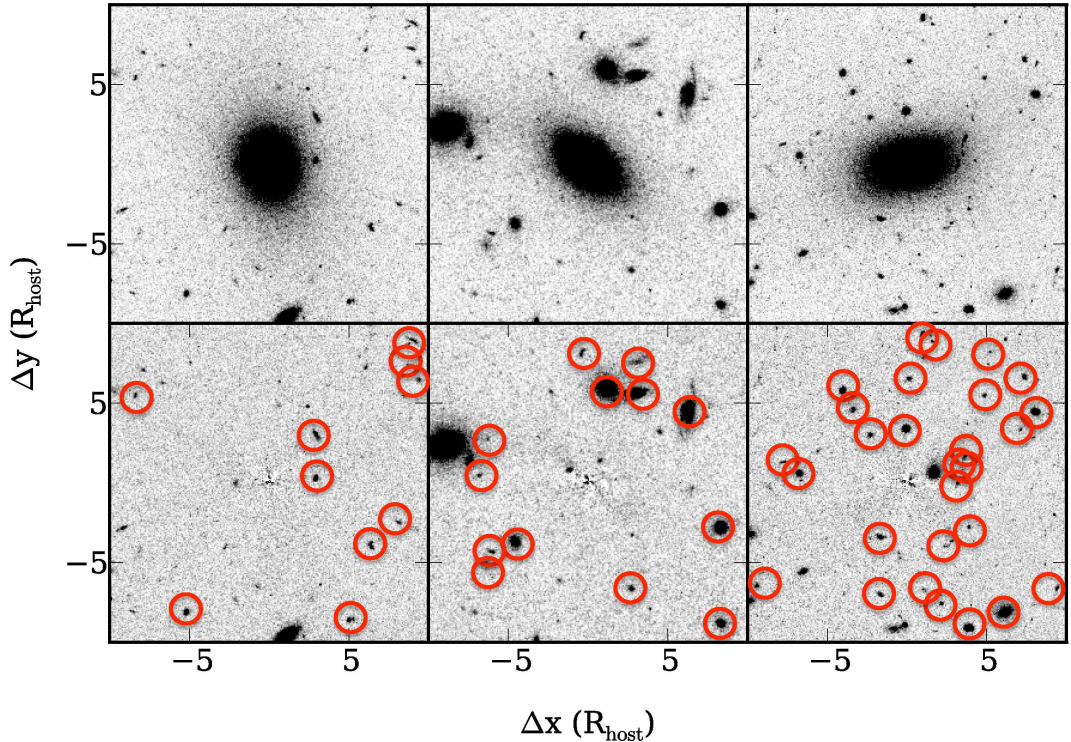}
 \caption{Demonstration of our image processing technique. The upper
 panels show the original Hubble Space Telescope images centered on
 main galaxies, while lower panels show the host-subtracted
 images. Red circles show objects we detect near the hosts. While we
 study the satellite population out to much larger radii (about 5
 times larger than the images), the host subtraction in the central
 region is essential for allowing an accurate characterization of the
 satellite spatial distribution, which in turn allows us to isolate
 the properties of the satellite population. Some objects that are too
 close to the edges, or which are below our detection threshold are
 not circled. Objects very close to the centers of the host galaxies
 are excluded as the host subtraction becomes inaccurate in the inner
 regions.}
\end{center}
\label{fig:hst}
\end{figure*}

For convenience of the reader, the detection technique is illustrated in
Figure~\ref{fig:hst}. First, the depth and high angular resolution of
Hubble images was used to identify satellites as much as a thousand
times fainter than their host galaxies, and out to redshift $z=0.8$. 
After adding simulated satellites near bright central galaxies, and testing object recovery and accuracy, 
the analysis was restricted to objects with F814$<25$ mag AB, in order to ensure completeness.

Using a single band of photometry, \citet{Nierenberg++12} model the
number density of objects near the chosen host galaxies as a
combination of a uniform density of background/foreground objects, in
addition to a population of satellite galaxies which have a number
density which increases radially as a power law near the host galaxies
\citep[e.g.][]{Chen++08, Watson++11,
Nierenberg++11,Nierenberg++12}. The power of the radial distribution
as well as the number of satellites within a fixed magnitude offset
from the host magnitude are simultaneously inferred to yield a
cumulative luminosity function.

\citet{Nierenberg++12} showed that the inferred satellite luminosity
function for low mass host galaxies is fully consistent with the
luminosity function of Milky Way satellites, and the satellites of
other similar mass galaxies at low redshift \citep{Strigari++12,
Liu++11, Guo++11}. Taking into account the differences between global
background subtraction rather than the locally estimated background
used in our measurement, which can cause a factor of two increase in
the inferred number of satellites due to correlated structure around
massive host galaxies, our results are also consistent with
\citet{Wang++12} at low redshifts and for the most massive satellites
($\log[L_{s}/L_{h}]> -1$), our results are consistent with those of
\citep{Newman++11} at higher redshifts.

 As we will show below, these new constraints on host mass and
redshift dependence of the satellite luminosity function provide
distinguishing power between the physical effects of varying the halo
mass function and star formation physics on the satellite luminosity
function.

\section{Theoretical Models}
\label{sec:theory}
We study three different semi-analytic models implemented in three different cosmological Cold Dark Matter simulations. In addition, we apply one of the semi-analytic models to a Warm Dark Matter cosmology. This section provides a description of both the Cold and Warm Dark Matter simulations and their models for galaxy formation.

\subsection{Cold Dark Matter Models}
When studying the satellite luminosity function, it is extremely useful to compare multiple cosmological simulations along with different star formation parameters in order to understand the range of theoretical predictions given different assumptions. In this work we focus on three Cold Dark Matter cosmological simulations, each with its own semi-analytic model for galaxy. 
The simulation parameters governing star formation and the underlying dark matter mass function were fully specified prior to performing this test and are the same as in previous papers, in which a more complete description can be found \citep{Menci++12, Lu++12, GuoQi++11}.  Below we briefly summarize the relevant aspects of these simulations.

\subsubsection{Menci et al. 2012}
The semi-analytic models are described in detail by \citet{Menci++12}. 
In brief, the backbone of the models are computations of dark matter 
merging histories that can resolve halos down to masses of 
$10^5 $M$_\odot$, allowing for an extremely accurate characterization of the survival and merger
histories of satellites. This is essential for studying the faint end
of the satellite luminosity function. The models predict mass
distributions of halos, and the merging rates of DM halos and subhalos
based on an input power spectrum, and generate
luminosities for these halos based on a set of star formation
prescriptions. 

These galaxy formation prescriptions connect the complex baryon physics of star formation to the 
dynamical evolution of the DM haloes. For each subhalo hosting a galaxy, the model predicts
radiative gas cooling, ensuing star formation and
associated feedback from Supernovae (SNae) events.
Cooled gas settles into a rotationally supported disk with radius and rotational velocity
related to the DM mass of the subhalo. This gas gradually condenses  into
stars at a rate consistent with the observed Kennicut-Schmidt law \citep{Kennicutt++98}.
SNae return part of the cooled gas to the hot gas phase
at the virial temperature of the halo. Star formation is also triggered by 
galaxy-galaxy interaction-driven starbursts, which induce gas accretion onto 
the central supermassive Black Holes. 

It is important to note that the set of parameters used in these models is much smaller than the 
number of observables they are simultaneously consistent with. In brief, the star formation parameters are tuned
to simultaneously match the field luminosity function in multiple bands from redshifts 0 to 6, the Tully-Fisher
relation \citep{Tully++77}, the stellar mass function of field galaxies from between redshifts 0 and 4, the stellar mass-star formation 
relation to a redshift of 2, the colors and color magnitude relations of both field and cluster galaxies in addition to the AGN luminosity function. 

\citet{Nierenberg++12} required host galaxies to be outside of R$_{200}$ of any galaxy with a higher stellar mass, where R$_{200}$ was estimated using the
formula from \citet{Dutton++10}. Ideally this selection would be duplicated when comparing with the simulation, however this is not possible
for the \citet{Menci++12} models, as they do not contain spatial information. This should not significantly affect our results, as comparisons 
with the Millennium and Bolshoi N-Body simulations (described below) show that the \citet{Nierenberg++12} central isolation criteria are efficient at returning central host galaxies ($\sim 90 $\%), and that the use of the matching selection
criteria in these cases did not significantly alter the inferred predicted number of satellites per host, relative to simply using the simulation selection of central galaxies.

\subsubsection{Guo et al. 2011}
The semi-analytic model of \citet{GuoQi++11} is based on two very large dark matter simulations, the Millennium Simulation \citep[MS;][]{Springel++05} and the Millennium-II Simulation \citep[MS-II;][]{Boylan-Kolchin++09}. The box size of the MS is $500h^{-1}\rm{Mpc}$ and its merger trees are complete for subhaloes above a mass limit of $1.7\times10^{10}h^{-1}\rm{M}_{\sun}$. The MS-II follows a cube of side $100h^{-1}\rm{Mpc}$, but with 125 times better mass resolution than the MS (subhalo masses greater than $1.4\times10^{8}h^{-1}\rm{M}_{\sun}$). Both simulations adopt the WMAP1-based $\Lambda$CDM cosmology \citep{Spergel++03} with parameters $h=0.73, \Omega_m=0.25, \Omega_\Lambda=0.75, n=1$ and $\sigma_8=0.9$. Due to the lower resolution of the MS, \cite{Wang++12} found that the luminosity functions of satellites flattens for satellites with $M_r > -18$. For brighter satellites with $M_r < -18$, the simulations are consistent. Since satellite luminosity functions will be measured to about three-orders of magnitude fainter than central primaries in this paper, we will focus on the semi-analytic model implemented on MS-II hereafter.

In general, the galaxy evolution model of \citet{GuoQi++11} is based on those developed by \citet{Springel++05,Cro++06b, D+B07}. The model includes a few main modifications such as the different definition of satellite galaxies, the gradual stripping and disruption of satellites, a mass-dependent model of supernova feedback, a modified model for reionization and a more realistic treatment of the growth of stellar and gaseous disks. Free parameters of these models by \citep{GuoQi++11} were determined to give close predictions to the abundance and clustering of low redshift galaxies, as inferred from SDSS, and are functions of their stellar mass, luminosity and color.

 There are two types of satellites in the simulation: $i)$ those with an associated dark matter subhalo (type-1) and $ii)$ those whose dark subhalo has fallen below the resolution limit of the simulation (type-2). For the latter, the position and velocity of the orphan galaxy is given by those of their most bound particle. Type-2 satellites are removed from the galaxy catalogues when one of these two conditions is fulfilled: $1)$ the time passed from the disruption of the subhalo is longer than their estimated dynamical-friction timescale, or $2)$ the integrated tidal forces from the host halo exceed the binding energy of the galaxy.

Here we use the data downloaded from http://www.mpagarching.mpg.de/millennium for registered users. 
We project the simulation box in three orthogonal directions (along their $x$, $y$ and $z$ axes).  In each projection we assign each galaxy a redshift based on its ``line-of-sight'' distance and peculiar velocity. We select isolated primaries using criteria which are directly analogous to those used by \citet{Nierenberg++12}. Satellites are defined to be all companion galaxies whose distances to the isolated primaries are smaller than the halo virial radius (R$_{200}$).  To directly compare results based on COSMOS, the real R$_{ \rm vir}$ provided in the database is ignored, and instead we used the empirical formula of \cite{Dutton++10} to estimate M$_{\rm vir}$ and R$_{200}$ from stellar masses of galaxies.

\subsubsection{Lu et al. 2012}

The baryonic processes implemented in this semi-analytic model are described in \citet{Lu++11,Lu++12}. Different from the previously published versions, this version of the model is applied on a set of halo merger trees extracted from a large cosmological $N$-body simulation, the Bolshoi simulation \citep{Klypin++11}, which has a box size 250$h^{-1}$ Mpc  on a side. The simulation adopted a cosmology favored by WMAP7 data \citep{Jarosik++11} and WMAP5 data \citep{Dunkley++09, Komatsu++09} with parameters with $\Omega_{\rm m,o} = 0.27$, $\Omega_{\Lambda,o} = 0.73$, $\Omega_{\rm b,o} = 0.044$, $h = 0.70$, $n = 0.95$ and $\sigma_8 = 0.82$. 
The mass resolution of the simulation is $1.35\times10^8 h^{-1}$M$_\odot$, which allows us to track halos and subhalos with mass $\sim 7\times 10^9  h^{-1}$M$_\odot$. Dark matter halos and subhalos are identified with the Rockstar halo finder \citep{Behroozi++12} based on adaptive hierarchical refinement in phase-space. 

As with the two other models in this work, the semi-analytic model follows the dark matter merger tree and calculates the rates of gas cooling, star formation, outflow induced by star formation feedback, and galaxy-galaxy mergers. The kinematics of satellite galaxies is followed by using subhalo information from the simulation whenever the subhalo is resolved. 
When a halo becomes a subhalo, we instantaneously strip the hot gas associated with the halo, while the stellar mass and cold gas mass remain intact. 

When the subhalo is no longer resolved in the simulation, the model applies dynamical friction \citep{Binney++87} to estimate when the satellite has merged into the central galaxy, and it assumes that the tidal stripping is strong enough to also strip the stellar mass and cold gas mass. At this point, the entire cold gas disk is stripped and is mixed into the hot gas of the host primary halo. Starting from that time, a fraction of stellar mass is tidally stripped per orbital timescale. The efficiency of the tidal stripping is controlled by a parameter, which is tuned to yield 30\% of the stellar mass is stripped in every orbital timescale to match the conditional stellar mass function of local galaxies. 

Other parameters governing star formation and feedback are tuned using an MCMC optimization to match the local galaxy stellar mass function (Moustakas et al 2013). The model is guaranteed to produce a global galaxy stellar mass function which provides the best possible match to the data between $10^9$ and $10^{12}$ M$_\odot $ at redshift zero, within the observational uncertainty and given the chosen model parametrization. 

To compare with observation, hosts were selected using the same isolation criteria as used in \citet{Nierenberg++12}, with hosts required to be not within R$_{200}$ of a host with higher stellar mass, where R$_{200}$ is estimated from \citet{Dutton++10}, using the relationship for early-type hosts. The satellite luminosity function was measured within this region, rather within R$_{\rm vir}$ as given by the simulation.

 \subsection{Menci Warm Dark Matter Model}
\label{sec:WDM}
 Warm Dark Matter has been proposed as a means of suppressing the
satellite luminosity function by reducing the number of low mass dark
matter halos, suppressing power with respect to CDM below a certain
cutoff-scale.  

In this work, we use WDM merger trees from
\citet{Menci++12} which are based on a cutoff scale of $\sim1$ Mpc,
corresponding to a $\sim$0.75 keV/c$^2$ thermal relic, which
suppresses the power spectrum on sub-Mpc scales and has behavior
equivalent to that of a $\sim$ 3.4 keV/c$^2$ sterile neutrino. We
chose this mass to be low enough to affect satellite-scale structure
while still agreeing with limits from observations of large-scale and
local group structure which constrain the particle mass to be larger
than 0.6 keV/$c^2$ for a thermal relic and 2.5 keV/c$^2$ for a sterile
neutrino DM particle \citep{Viel++09, Boyarsky++09, Polisensky++11, Kang++12}.

The fiducial mass function from \citet{Menci++12} was based on an Extended Press Schechter formalism \citep[EPS,][]{Bond++91}. In our predictions, we consider the effects of a complete suppression of progenitors with masses below that corresponding to the free-streaming scale; this  maximizes the possible effects of different window functions and collapse thresholds in the WDM merging trees \citep[see][]{Benson++13}. We include the possible range of model predictions under different assumptions for the building up of merging trees in the uncertainty regions in Figure \ref{fig:comparison}. 

The mass function from \citet{Menci++12} is based on on the Extended Press
Schechter formalism \citep[EPS,][]{Bond++91}, which is modified to take
into account the suppression of structure below the free-streaming
scale \citep[see, e.g.][]{Menci++12,Benson++13}. The uncertainties
associated with this modification are reflected in the width of the
prediction for the final WDM satellite luminosity function.

 Our goal in this comparison was to explore how the effects of varying the power spectrum
compare with the effects of varying semi-analytic
star formation parameters. To achieve this,  \emph{the same} semi-analytic model of star
formation is used for the WDM models as was applied to the CDM merging trees in the Menci
CDM model.  The CDM semi-analytic model was selected to provide a good match to the field 
color distribution. \citet{Menci++12} show that this model provides a good fit to the field luminosity function
when applied to a WDM power spectrum.

\section{Results}
\label{sec:results}

The improvements in observations and the implementation of
semi-analytic models in cosmological scale simulations allows a
comparison between the observed and predicted number of satellites as
a function of host galaxy mass and cosmic time for the first time,
thus allowing for significantly more discriminatory power than tests
based only on the MW or the local volume.

\begin{figure*}
\begin{center}
\includegraphics[width=1.\textwidth,clip]{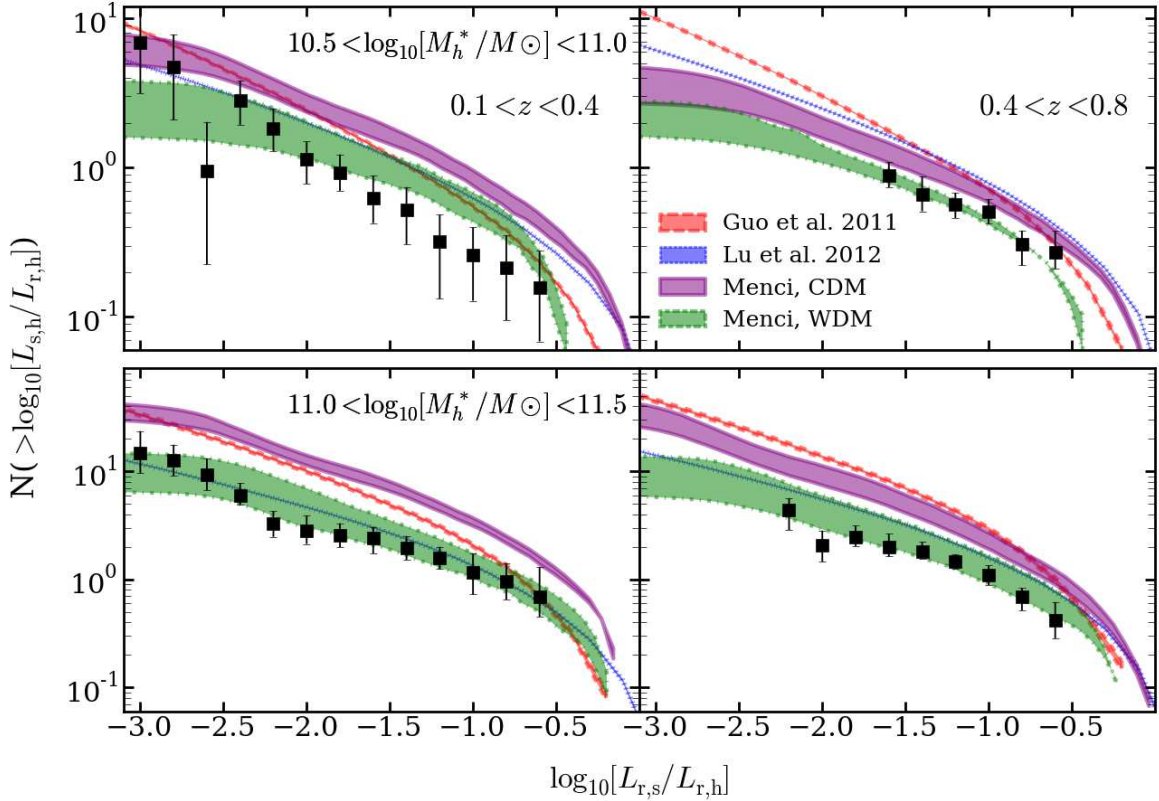}
 \caption{Comparison of CDM and WDM model predictions to observed number of
 satellites over a range of redshift and host stellar masses.
 The purple, blue and red curves represent predictions from three CDM based simulations with separate semi-analytic models for star formation. The gray curve shows the prediction for one WDM simulation with star formation parameters identical to those used in Menci, CDM prediction. For the Lu and Guo models, the line widths represent the scatter in the mean value across the simulation boxes. For the Menci CDM model, the shaded region represents the range of model predictions possible given the observational uncertainty in the selection of host halo masses. For the Menci WDM model, the shaded region also accounts for the uncertainties in the suppression of progenitors below the free-streaming scale (see \ref{sec:WDM}) The points with
 vertical error bars are measurements \citep{Nierenberg++12}. The top panels show the comparison for lower mass hosts at lower (left) and higher redshift (right), while the lower panels are the equivalent for higher mass hosts.}
\label{fig:comparison}
\end{center}
\end{figure*}

The results of the comparison are presented in
Figure~\ref{fig:comparison}, where we plot the observed number of
satellites as a function of the ratio between host and satellite
luminosities in SDSS-r, along with the WDM and CDM model predictions. Among the CDM model predictions, two main trends are evident. First, no one CDM prediction precisely matches the observation at all redshift and stellar mass intervals. This highlights the importance of using data from a range of redshift and stellar mass when tuning the parameters of semi-analytic models. All three models show qualitatively similar behavior with a strong dependence on the number of satellites per host on the host galaxy mass. Taking into account the covariance between data points
(calculated by bootstrap resampling), the generalized chi-squared
between the models and the data is 529, 653 and 105 for the Menci, Guo and Lu models respectively for 41 degrees of freedom.  

The Menci WDM model shows distinct behavior in comparison with the CDM models. Notably, it predicts weaker host mass dependence and less redshift evolution than predicted by any of the CDM models. Of the four models it provides the best agreement with the data with a chi-squared of 56.  
 
 One of the most interesting results of the above comparison is the significant difference in the predicted satellite luminosity functions, even among the three CDM models. In the following section, we perform a detailed comparison of some of the properties of the four models in order to explore the cause of these differences.

\section{Comparison of simulation properties}
\label{sec:comparison of sims}
In the previous section we found significant differences in the model predictions for the satellite luminosity function. As discussed in Section \ref{sec:theory}, numerous physical processes contribute to the final predicted luminosity function. In this section, we compare key aspects of the models in order to elucidate which model assumptions drive the predicted differences. We first compare the subhalo mass function in the models, before the effects of tidal stripping take effect, for all subhalos regardless of final luminosity. We then show the stellar mass to halo mass relationship for hosts and satellites in the luminosity range considered in this work, before the subhalos have undergone tidal stripping. We conclude by comparing the predicted satellite colors to show the differences in star formation histories.

\subsection{Unstripped subhalo mass function}
\begin{figure*}
\begin{center}
\includegraphics[width=1.\textwidth,clip]{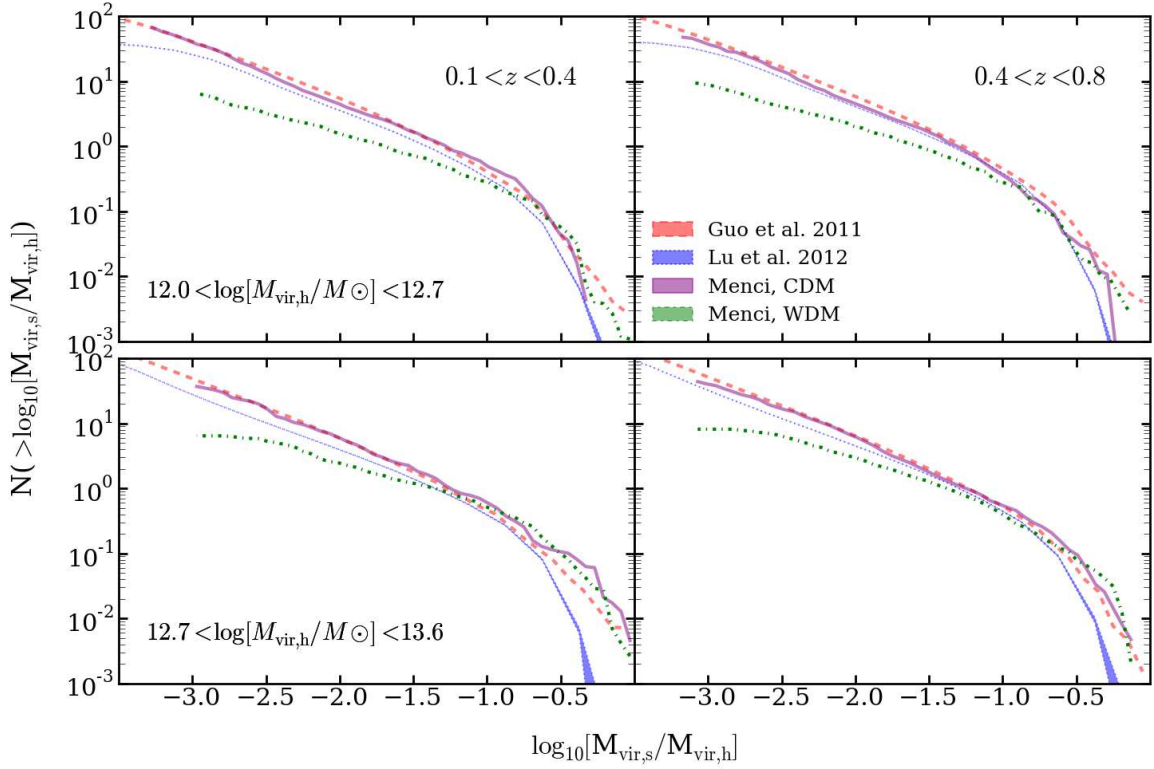}
\caption{Comparison of the subhalo mass function at the time of subhalo accretion for the four models, for a range of masses representative of the host halos and subhalos in Figure \ref{fig:comparison}. Note that these mass functions do not include the effects of tidal stripping by the main halo, nor are they selected to host luminous satellite galaxies, or host galaxies with the same stellar masses as the hosts in Figure \ref{fig:comparison} }
\label{fig:dmMass}
\end{center}
\end{figure*}

The four simulations all assumed slightly different cosmologies and had different mass resolutions, so it is important to consider how much the differences in the predicted results are due to these effects. In Figure \ref{fig:dmMass} we plot the mass function of subhalos in a mass range representative of the hosts and subhaloes from Figure \ref{fig:comparison}. Note that this mass function is purely based on dark matter mass, and does not directly correspond to the satellite galaxies in Figure \ref{fig:comparison}, as there is no luminosity selection. 

The Guo, Menci, and Lu CDM subhalo mass functions are extremely similar, while the WDM model shows significant truncation for subhalos with $\log[M_{sub}/M_{host}]< -1$ for the less massive hosts and $\log[M_{sub}/M_{host}]< -1.5$ for the more massive hosts, and predicts similar behavior to other WDM models with sub-Mpc cutoff scales \citep[e.g.][]{Colin++00,Smith++11, Kamada++13}.

\subsection{The halo mass to stellar mass relationship}
One of the most important physical processes relied on in this comparison between simulation and observation, is the relationship between host halo mass and stellar mass, because the number of subhalos around a given host galaxy is strongly dependent on the halo mass of the host galaxy \citep[see e.g.][]{Busha++11}, particularly in the case of CDM. Furthermore, as we always consider the quantity $\log_{10}[L_{s}/L_{h}]> -3$, it is important to distinguish whether differences in the models are caused by differences in $L_{s}$ or in $L_{h}$. 

\begin{figure*}
\begin{center}
\includegraphics[width=1.\textwidth,clip]{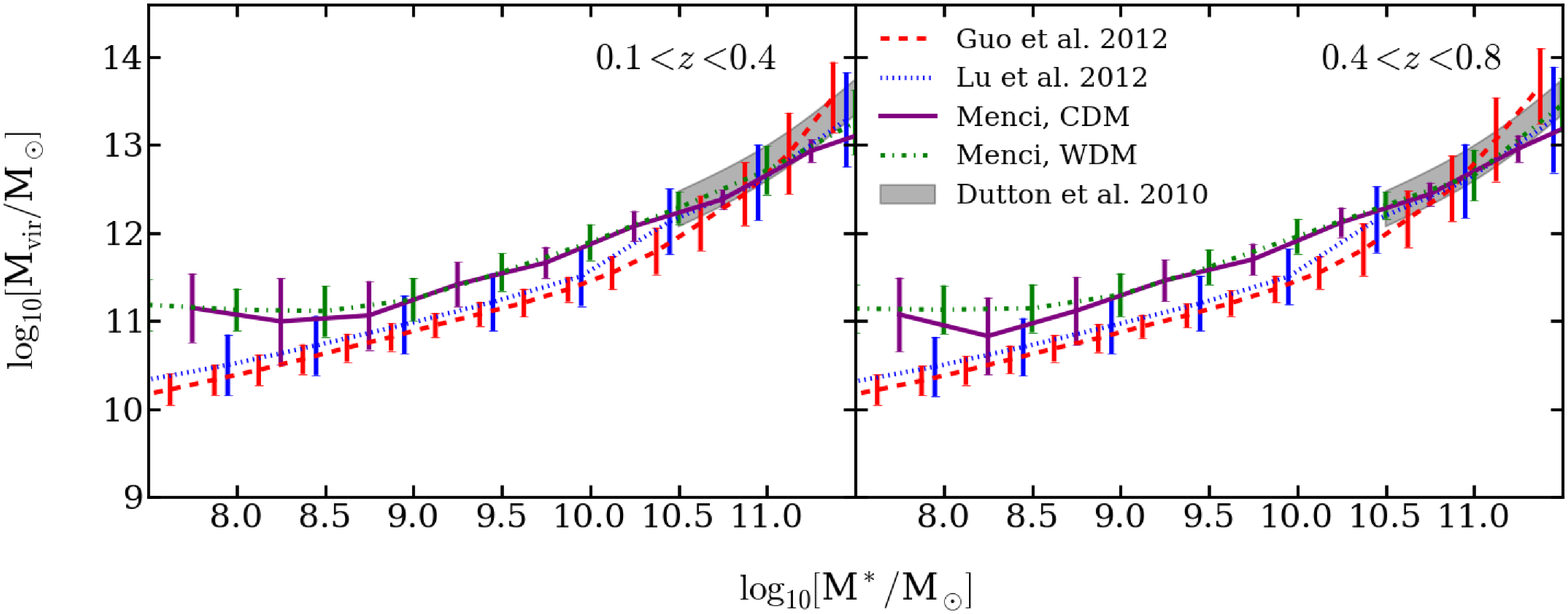}
\caption{ Comparison of the virial to stellar mass relationship for both host and satellite galaxies for the four models. The subhalo virial masses are taken at the time of accretion, and thus do not include the effects of tidal stripping. }
\label{fig:MhMsh}
\end{center}
\end{figure*}

In Figure \ref{fig:MhMsh} we plot the halo to stellar mass
relationship for the four models, in addition to the observed
relationship from \citet{Dutton++10}, which was used to estimate
R$_{200}$ for the host galaxies in the observations. For bright
galaxies with stellar masses greater than $10^{10.5}$M$_\odot$, the models are
all very similar to each other and to the observed relationship,
within the large scatter. Thus we conclude that the differences in
amplitude in the predictions for the satellite luminosity function are
not driven primarily by differences in the halo to stellar mass
relationship for host galaxies.  

Towards the faint end, all models are
very similar within the scatter, although the Menci models show a
marginally higher dark matter mass at fixed stellar mass.
As we show in the following section, the Menci satellite colors tend to be
bluer than the Guo and Lu colors. From \citep{Bell++03}, Table 7, the bluer colors
of the Menci model, correspond to an r band mass-to-light ratio
which is lower by roughly 0.1-0.2 dex on average than for the Guo and Lu models.
Thus, as the final luminosity functions are similar, it is expected that the Menci models
would produce galaxies with on average slightly lower stellar masses for fixed halo mass.

\subsection{Satellite galaxy colors}
\label{sec:colors}

\begin{figure*}
\begin{center}
\includegraphics[width=1.\textwidth,clip]{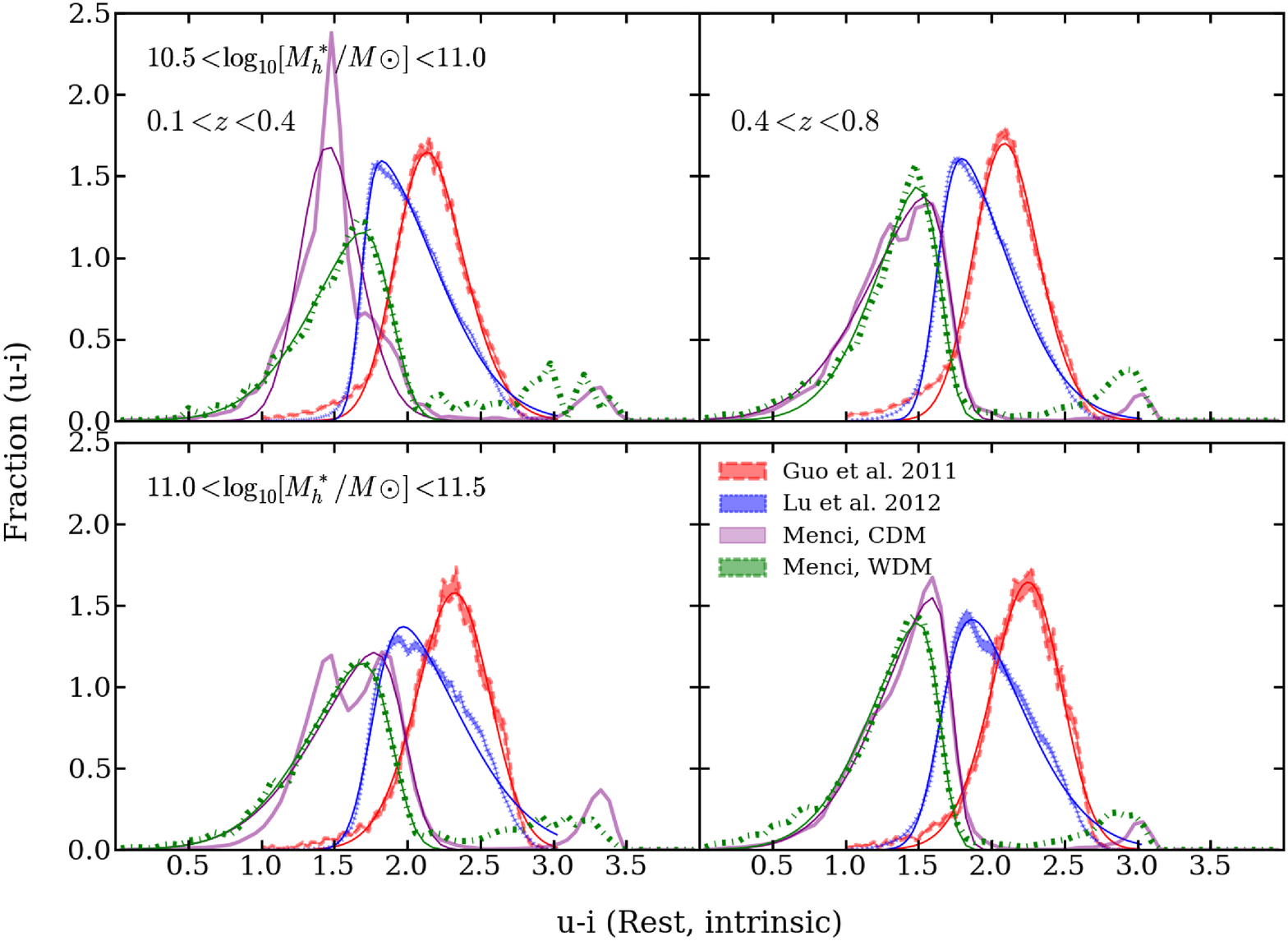}
\caption{The normalized rest-frame distribution of colors, for the satellite galaxies plotted in Figure 2. The solid lines indicate the best fit skewed normal distributions defined by Equation\ref{eq:skewed}, with fit parameters listed in Table \ref{tab:colorFit} }
\label{fig:colors}
\end{center}
\end{figure*}

The intrinsic colors of satellite galaxies are dependent on the star formation history. As discussed above, this is determined by the metallicity, feedback and UV heating, in addition to environmental effects as the satellites enter the influence of the host galaxy halo. All three models used the same stellar population synthesis models from \citep{Bruzual++03}, thus satellite galaxy colors provide an important means of distinguishing between different physical models for the suppression of star formation in low mass halos. Two models which produce similar luminosity functions may rely on very different star formation prescriptions, which will result in different color distributions.  In Figure \ref{fig:colors}, we plot the predicted distribution of rest-frame u-i colors for satellites with $\log_{10}[L_s/L_h]>-3$.

In order to summarize these predictions, we fit each of the distributions to a skewed normal distribution \citep{Azzalini++85} where the probability of having color $c$ is given by:
\be
P(c) \propto e^{\frac{(c-c_o)^2}{2\sigma^2}} \left(1+erf \left(\frac{a (c-c_o)}{\sqrt{2} \sigma} \right)\right)
\label{eq:skewed}
\ee

Here $c_o$ and $\sigma$ are the usual mean and standard deviations of a normal distribution while the parameter $a$ describes the skewness. The best fit values for these parameters are listed in Table \ref{tab:colorFit} \footnote{We do not consider the secondary peaks in the Menci models in this fit.}. The parametrization of the distributions as being skewed normal is not physically motivated, but rather intended to facilitate future comparison between observation and these predictions.

The models all predict significantly different color distributions for the satellite galaxies, with the exception of the Menci CDM and WDM models, for which the same star formation parameters were used.  We highlight the fact that although the luminosity function predicted by the Menci WDM and Lu CDM models are similar within some redshift and host stellar mass ranges, the color distributions are very different. This is due to the fact that in the Lu CDM model, the faint end of the luminosity function is suppressed mostly by the effects of feedback and ram pressure stripping and heating of the gas by the host halo, while  in the Menci WDM model, the luminosity function is suppressed by the lack of low mass subhalos, as expected.

\begin{figure*}
\begin{center}
\includegraphics[width=1.\textwidth,clip]{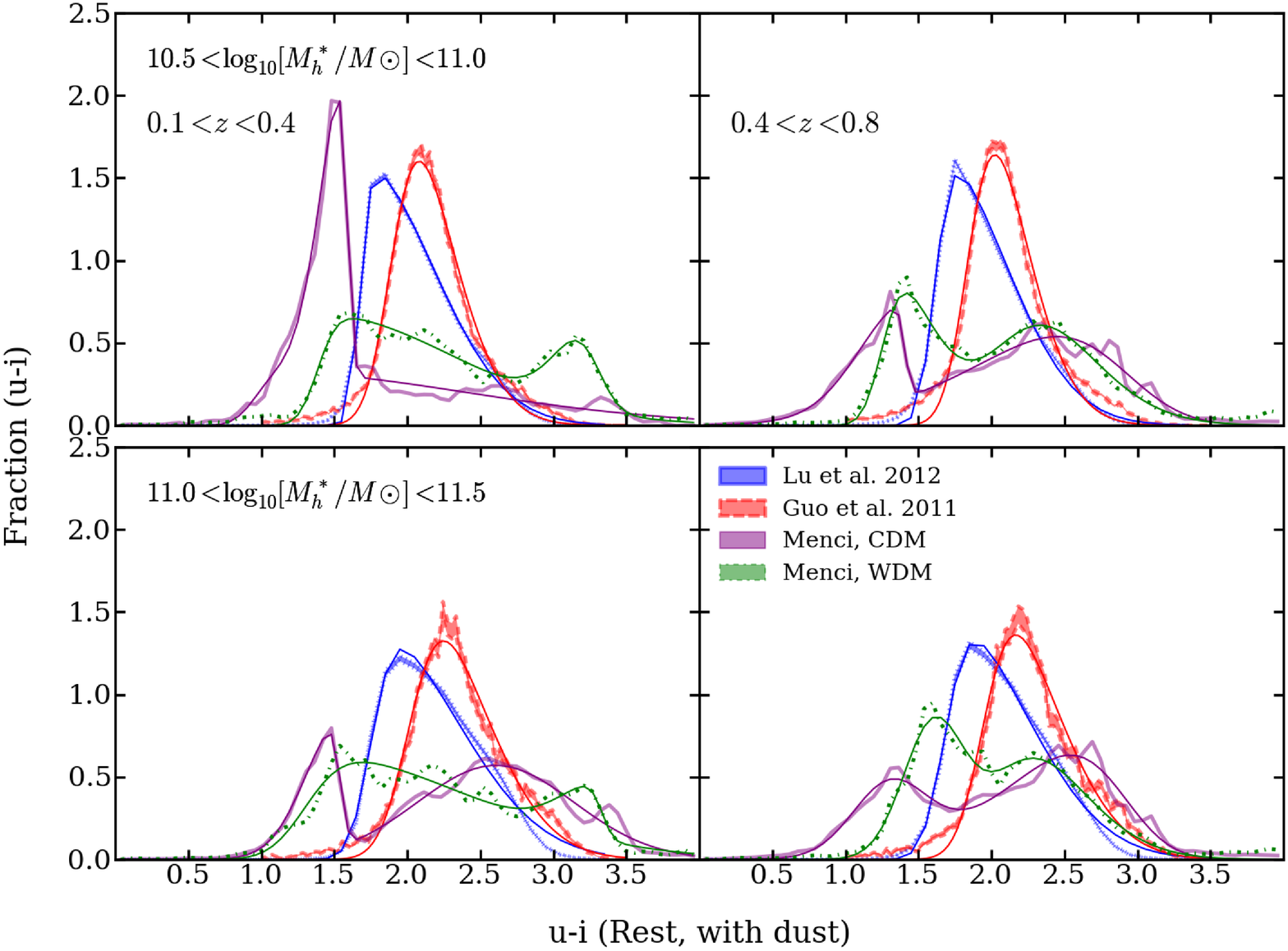}
\caption{The normalized rest-frame distribution of colors, with dust extinction for the satellite galaxies plotted in Figure 2, along with the best fitting skewed normal distributions. The Menci models are fit with a sum of two skew normal distributions, to account for the prominent secondary peak that appears.}
\label{fig:colors_dust}
\end{center}
\end{figure*}

In order to facilitate comparison with observation, it is important to also consider the effects of dust extinction. In Figure \ref{fig:colors_dust}, we show the prediction for the rest-frame colors of satellite galaxies with dust. The Menci models behave very differently with the addition of dust, relative to the Guo and Lu models, with the Menci color distributions becoming much wider and significantly redder while the Guo and Lu models develop a longer redward tail, without significant other alteration. The secondary peaks in the Menci model become more prominent with the addition of dust, so we fit the Menci models with the sum of two skewed normal distributions with parameters reported in Table \ref{tab:colorFit}, where the value $A_2/A_1$ describes the relative amplitudes between the two skewed normals.

\begin{deluxetable*}{llllll}
\tabletypesize{\small}
\tablecaption{\label{tab:colorFit} Best fit parameters of the color distributions}\\
\tablehead{\colhead{Host Properties} & \colhead{Dust}& \colhead{Guo et al. 2012}  & \colhead{Lu et al. 2012} & \colhead{Menci, CDM} &\colhead{Menci, WDM}}

\startdata

                                            &   &$c_o$, $\sigma$, $a$& $c_o$, $\sigma$, $a$  & $c_o$, $\sigma$, $a$,               &$c_o$, $\sigma$, $a$,  \\
                                            &   &                    &                       & $A2/A1$, $c_{o,2}$, $\sigma_2$, $a_2$ &$A2/A1$, $c_{o,2}$, $\sigma_2$, $a_2$  \\
                                             \tableline
                                                \noalign{\smallskip}

  $10.5<\log[\Mstar/\Msun$]$<11.0$, 0.1$<$z$<$0.4 & Yes     &1.7, 0.5, 8.8 & 1.9, 0.4, 2.5 & 1.6, 0.2, -5.0                      & 1.4, 1.0, 9.3                     \\
                                                  &         &              &               & 1.4, 1.0, 1.5, 12.5                 & 0.2, 3.3, 0.25, -1.9               \\
                                               
                                                                                                                                                                         \noalign{\smallskip}                        
                                                &No          &2.0, 0.3, 1.5 & 1.7, 0.5, 7.0  &1.3, 0.3, 1.4 & 1.9, 0.5, -4.0         \\ 
                                                \tableline
                                                \noalign{\smallskip}

$10.5<\log[\Mstar/\Msun$]$<11.0$, 0.4$<$z$<$0.8 & &1.6, 0.5, 5.6 & 1.8, 0.3, 2.4 & 1.4, 0.3, -9.0        & 1.3, 0.3, 3.0     \\
                                                & &              &               & 3.1, 2.9, 0.9, 2.5  & 1.5, 2.1, 0.5, 1.6 \\
                                               
                                                                                                                                                                         \noalign{\smallskip}                        
                                               &  &1.9, 0.3, 1.2 & 1.6, 0.4, 4.5  &1.7, 0.5,-5.4 & 1.7, 0.4, -4.4         \\ 
                                                 
                                               \tableline        
                                               \noalign{\smallskip}

$11.0<\log[\Mstar/\Msun$]$<11.5$, 0.1$<$z$<$0.4 &  &1.8, 0.6, 5.1 & 1.2, 0.5, 2.7 & 1.5, 0.24, -5.9 & 1.3, 1.1, 5.5    \\
                                                & &              &               & 4, 2.6, 0.5, 0 &  0.3, 3.3, 0.3,-4.5\\
                                               
                                                                                                                                                                         \noalign{\smallskip}                        
                                                & &2.5, 0.3, -1.2 & 1.8, 0.5, 4.3  &2.0, 0.5,-4.4 & 1.9, 0.5, -4.2         \\ 

                                                \tableline
                                                \noalign{\smallskip}

$11.0<\log[\Mstar/\Msun$]$<11.5$, 0.4$<$z$<$0.8& &1.7, 0.5, 4.6 & 1.9, 0.5, 2.9 & 1.3, 0.25, 0        & 1.6, 0.2, 0        \\
                                               &  &              &               & 2.3, 2.9, 0.7, -2.6 &  1.3, 2.0, 0.5, 1.7 \\

                                                                                                                                                                         \noalign{\smallskip}                        
                                               &  &2.4, 0.3, -1.3 & 1.7, 0.5, 3.7  &1.7, 0.5,-6.2 & 1.7, 0.4, -5.3

\enddata
\end{deluxetable*}

\section{Comparison with literature}
\label{sec:literature}
The models presented in this work predict a broad range of observables, many of which have already been compared to observations, either at lower redshifts, or of brighter objects than those considered in this paper. Below we provide a brief comparison with some of these observations, a more detailed discussion of many of these comparisons can be found in the original papers describing the models \citep{Menci++12, GuoQi++11, Lu++12}.

\subsection{Satellite luminosity function}
Studies of the satellite luminosity function typically focus on satellite galaxies at low redshifts with $z<0.1$, both in simulation and observations. As we have shown in \citet{Nierenberg++12}, the COSMOS field can be used to study satellites at higher redshifts, and yields luminosity functions at low redshifts which are consistent with satellite luminosity functions from SDSS \citep[e.g.][]{Guo++11, Liu++11,Lares++11,Strigari++12}, and the Milky Way satellite luminosity function \citep{Toll++08}. All four of the models provide good fits to the satellite luminosity function of Milky Way mass hosts at low redshifts, by design. The extra information of redshift and host galaxy stellar mass provides additional constraints to these models. For instance, when comparing to bright field galaxies observed by \citet{Perez++08}, \citet{GuoQi++11} found their model agreed well at low redshift, while becoming discrepant with observation by a redshift of 1.

\subsection{Satellite Colors}
In Section \ref{sec:colors} we showed that the three CDM models predicted significantly different distributions for the colors of faint satellite galaxies. Although we cannot directly test this prediction in this work, some comparisons can be made with measurements of field galaxies at low redshifts. In particular, \citet{GuoQi++11} found that their model predicted colors matched observed SDSS colors well except for low masses ($\log_{10}[M^*/M_{\odot}]<9.5$), which were redder in the simulation than in observations. \citet{Menci++12} found that down to an absolute magnitude of $M_r = -18$ (roughly a stellar mass of $\sim $9 $ \log[M^*/M_{\odot}]$), their color distribution agreed well with SDSS measurements from \citet{Baldry++04}. They did not compare for fainter satellites.  

The Lu and Guo semi-analytic models were not optimized to reproduce the observed galaxy color-magnitude relation at low redshifts, thus future implementations of these semi-analytic models which include this information may yield significantly different predictions for the color distribution of satellite galaxies.
 
Recently, \citet{Knobel++12} measured the red fraction of massive ($\log_{10}[M^*/M_{\odot}]>10$) satellites between redshift 0.1 and 0.8, and found no evidence for significant evolution over this time, indicating that these low redshift results may apply to higher redshifts. From a different point of view,  \citet{Behroozi++12} used abundance matching techniques to infer that low stellar mass ($\log_{10}[M^*/M_{\odot}]\sim9$)  \emph{field} galaxies on average continue to form a significant fraction of their stars at redshift $\sim 1$. However, this result cannot be directly applied to satellite galaxies, as star formation in satellite galaxies has been shown to be quenched relative to field galaxies at low redshifts \citep{Pasquali++10, Geha++12, Kauffmann++04}.

 Future measurements of faint satellite colors since intermediate redshifts will provide interesting new constraints on these models, by helping to distinguish between suppression of the satellite luminosity function by environmental quenching and supernovae feedback, or the subhalo mass function at the low mass end.

\section{Discussion and conclusion}
\label{sec:discussion}
We have compared predictions from three current semi-analytic models
applied to CDM simulations and one semi-analytic model applied to a
WDM simulation to observations of satellite galaxies in two redshift
intervals and with two host mass bins. By comparing multiple models,
we demonstrated that the effects of varying star-formation
prescriptions within semi-analytic models, and varying the underlying
dark matter mass function have fundamentally different effects on the
predicted host stellar mass dependence, and redshift evolution of the
satellite luminosity function. In particular, the WDM
model predicts a satellite luminosity function with much weaker host
stellar mass dependence and redshift evolution than any of the CDM
models.

We find that the WDM provides the best match to observation in all
redshift and host mass intervals, most closely matching the host mass
dependence and lack of redshift evolution in the data.  This
comparison highlights the importance of comparing models of satellite
galaxy evolution to observations from a range of redshifts and stellar
mass regimes, as a model that provides a close match to observation in
one regime can perform more poorly than other models in a different
regime.

This exploration of different CDM models suggests that {\it current}
models generically have difficulty reproducing the mass and redshift
dependence of the satellite luminosity function. However, future
improvements to the semi-analytic models for star formation
prescriptions may allow for improvements between the CDM predictions
of the satellite luminosity function.  In fact, as shown in the paper
by \citet{Nierenberg++12} subhalo abundance matching techniques
\citep{Busha++11} can reconcile our observations with the CDM subhalo
mass function. Although abundance matching is descriptive and not
directly linked to known physical processes, it is possible that
future semi-analytic models will be able to produce the stellar mass
to halo mass relation as illustrated by this approach.

Numerous studies have shown that abundance matching techniques can be used to map measured properties of star formation to simulated
halo masses, and thus provide important constraints for semi-analytic models \citep[e.g][]{Behroozi++12, Reddick++12, Behroozi++10, Busha++11}.
The descriptive power of abundance matching is limited for low mass satellite galaxies by the resolution of dark matter simulations and the depth
and redshift range of observations. Higher resolution simulations with varying dark matter power spectra, in conjunction 
with environmentally dependent abundance matching, can potentially provide very interesting constraints for the physical processes
governing star formation in low mass halos in varying cosmologies.

Additional observational data are needed to further constrain the
models and therefore help distinguish between whether the discrepancy between predicted and observed
satellite luminosity function can be mitigated by either
improved baryonic physics in CDM models or a WDM power spectrum.  For example, we
have shown in this work that the Lu CDM model which most closely
matches the data predicts a significantly different color distribution
for the satellite galaxies than the Menci WDM model, indicating that
future observations of the colors of faint satellites will provide an
important test of whether the luminosity function of satellites is
suppressed primarily by baryonic processes or by a WDM mass function.
We plan to carry out such measurement in the near future by exploiting
the rich multicolor datasets publicly available in the HST legacy
fields.


\acknowledgments

A.M.N. and T.T. acknowledge support from NSF through CAREER
grant. T.T. acknowledges support from the Packard Foundations through
a Packard Research Fellowship. We sincerely
thank A. Benson, A. Peters, R. Wechsler and
F. van den Bosch for useful comments and insightful
conversations. This work was based on observations made with the
NASA/ESA Hubble Space Telescope, and obtained from the Data Archive at
the Space Telescope Science Institute, which is operated by the
Association of Universities for Research in Astronomy, Inc., under
NASA contract NAS 5-26555.  These observations are associated with the
COSMOS and GOODS projects. Part of the work presented in this paper
was performed by AMN and TT while attending the program "First Galaxies
and Faint Dwarfs: Clues to the Small Scale Structure of Cold Dark
Matter" at the Kavli Institute of Theoretical Physics at the
University of California Santa Barbara, supported in part by the
National Science Foundation under Grant No. NSF PHY05-51164.

\bibliographystyle{apj}
\bibliography{references}


\end{document}